\documentclass[%
 aip,
 jcp,%
 amsmath,amssymb,
preprint,%
]{revtex4-1}

\usepackage{graphicx}% Include figure files
\usepackage{dcolumn}% Align table columns on decimal point
\usepackage{bm}% bold math
\usepackage{xcolor}
\usepackage{bbold}
\begin{document}

\author{Jakub K. Sowa}
\email{jakub.sowa@oxon.org}
\affiliation
{Department of Chemistry, Rice University, Houston, Texas 77005, USA}
%\altaffiliation{Center for Adapting Flaws into Features, Rice University, Houston, Texas 77005, USA}
\author{Peter J. Rossky}
\email{peter.rossky@rice.edu}
\affiliation
{Department of Chemistry, Rice University, Houston, Texas 77005, USA}
%\altaffiliation{Center for Adapting Flaws into Features, Rice University, Houston, Texas 77005, USA}

%%%%%%%%%%%%%%%%%%%%%%%%%%%%%%%%%%%%%%%%%%%%%%%%%%%%%%%%%%%%%%%%%%%%%
%% The document title should be given as usual. Some journals require
%% a running title from the author: this should be supplied as an
%% optional argument to \title.
%%%%%%%%%%%%%%%%%%%%%%%%%%%%%%%%%%%%%%%%%%%%%%%%%%%%%%%%%%%%%%%%%%%%%
\title
  {A Bond-Based Machine Learning Model for Molecular Polarizabilities and A Priori Raman Spectra}

\begin{abstract}
 The use of machine learning (ML) algorithms in molecular simulations has become commonplace in recent years.
There now exists, for instance, a multitude of ML force field algorithms that have enabled simulations approaching \textit{ab initio} level accuracy at time scales and system sizes that significantly exceed what is otherwise possible with traditional methods.
Far fewer algorithms exist for predicting rotationally equivariant, tensorial properties such as the electric polarizability.
Here, we introduce a kernel ridge regression algorithm for machine learning of the polarizability tensor. This algorithm is based on the bond polarizability model and allows prediction of the tensor components at the cost similar to that of scalar quantities.
We subsequently show the utility of this algorithm by simulating gas phase Raman spectra of biphenyl and malonaldehyde using classical molecular dynamics simulations of these systems performed with the recently developed MACE-OFF23 potential. The calculated spectra are shown to agree very well with the experiments and thus confirm the expediency of our algorithm as well as the accuracy of the used force field. 
More generally, this work demonstrates the potential of physics-informed approaches to yield simple yet effective machine learning algorithms for molecular properties.
\end{abstract}

\maketitle
%%%%%%%%%%%%%%%%%%%%%%%%%%%%%%%%%%%%%%%%%%%%%%%%%%%%%%%%%%%%%%%%%%%%%
%% Start the main part of the manuscript here.
%%%%%%%%%%%%%%%%%%%%%%%%%%%%%%%%%%%%%%%%%%%%%%%%%%%%%%%%%%%%%%%%%%%%%
\section{Introduction}
Machine learning (ML) algorithms are now well established tools in molecular simulations.\cite{noe2020machine,deringer2021gaussian} 
Most efforts to date have focused on the development of ML force fields, trained on potential energies and their gradients, that can subsequently be used in molecular dynamics (MD) or Monte Carlo simulations.\cite{unke2021machine,behler2007generalized,behler2016perspective,poltavsky2021machine,deringer2019machine,doerr2021torchmd} Such force fields have transformed the field of molecular simulations by enabling approximately \textit{ab initio}-level calculations at a fraction of the cost and on much larger system sizes than possible with conventional \textit{ab initio} methods.
These algorithms have been especially impactful in the case of materials for which no well-established analytical force fields exist as well as in `reactive' systems undergoing bond formation and breaking events.
The former include a range of inorganic and hybrid inorganic-organic materials, from amorphous carbon\cite{deringer2017machine} to semiconducting nanocrystals passivated with organic ligands,\cite{sowa2023exploring} while the latter group comprises for instance many catalytic systems as well as compounds undergoing proton transfer reactions.\cite{andrade2020free,schaaf2023accurate,kaser2020reactive,houston2024formic}

Apart from ML force fields, more recently, models for machine learning tensorial properties such as molecular dipole moments and polarizability tensors have also been developed.\cite{gastegger2017machine,veit2020predicting,grisafi2018symmetry,wen2022deep,schienbein2023spectroscopy} 
Crucially, together with ML force fields, such algorithms enable calculations of the IR and Raman spectra, respectively.\cite{sowa2024ir,raimbault2019using,sommers2020raman,grumet2024delta} This is highly valuable not only as an aid in the interpretation of the measured signal but also as a direct way to benchmark the performance of ML algorithms against experiment.\cite{fedik2022extending} 

Very broadly, most ML algorithms used in molecular simulations can be divided into two categories: those based on kernel (and related Gaussian Process Regression) algorithms\cite{deringer2021gaussian,christensen2020fchl,chmiela2017machine,rupp2012fast,grisafi2018symmetry} and those based on (almost invariably deep) neural networks.\cite{behler2007generalized,schutt2018schnet,smith2017ani,zhang2018deep,doerr2021torchmd} 
These two classes of methods are in many ways complementary.  
The kernel methods often require less data, as compared to neural net algorithms, to achieve the desired accuracy. They also typically posses few hyper-parameters and their training is non-iterative. By contrast,  the architecture of the neural network as well as the training procedure need to be, at least in principle, tuned to ensure optimal performance. The major disadvantage of the kernel-based algorithms is that, unlike in the case of neural-net machine learning, the training data become de facto parameters of the model resulting in algorithms whose computational cost increases with the size of the training set. Consequently, kernel-based methods are particularly well-suited to relatively small or homogeneous systems.

The translational and rotational invariance of scalar properties, such as the potential energy, in machine learning algorithms is typically ensured simply by converting molecular geometries into translationally and rotationally invariant descriptors.\cite{rupp2012fast,behler2007generalized,behler2011atom} 
Developing ML models for rotationally equivariant tensorial quantities on the other hand generally constitutes a more challenging task.
For the electric polarizability, a rank-2 tensor and a focus of this work, a number of neural network algorithms have been developed.\cite{gastegger2021machine,xu2024tensorial,sommers2020raman,batatia2022mace,nguyen2022predicting,feng2023accurate} Some of them rely on equivariant neural networks\cite{batatia2022mace,nguyen2022predicting} although physics-informed models based on the response formalism\cite{gastegger2021machine,zhang2023universal} as well as the Applequist’s dipole interaction model\cite{feng2023accurate} have also been proposed.
To the best of our knowledge, only two kernel-based methods for machine learning of the polarizability tensor exist in the literature. The first of these\cite{liang2017solvent,raimbault2019using,martinka2024simple} relies on rotating and aligning all the considered structures, and treating the individual tensor components as scalar quantities. Such an approach however generally requires much larger amounts of training data in the case of flexible or reactive molecular systems.
The more sophisticated symmetry-adapted Gaussian process regression method of Grisafi \textit{et al.}\cite{grisafi2018symmetry,wilkins2019accurate} involves a generalisation of the scalar kernel ridge regression (KRR) to tensorial properties by replacing the scalar kernel with one averaged over rotational symmetry operations.
%
%a rotationally equivariant form through an integration of the kernel over rotational operations.

In this work, we introduce a simple alternative to these methods. Our approach can be viewed as a KRR implementation of the bond polarizability model (BPM)\cite{cardona1982light,wolkenstein} which assumes that the overall molecular polarizability is a sum over bond contributions. As we will demonstrate, constructing an algorithm in such a way results in a simple yet effective polarizability ML model.
This work is structured as follows.
We begin in Section \ref{dology} by briefly discussing the conventional bond polarizability model. We then introduce our kernel-based polarizability algorithm built on BPM, as well as outline the formalism used to obtain the Raman spectra. Section \ref{methods} provides an outline of the used computational methods. In Section \ref{results}, we apply our algorithm to two test cases: the biphenyl and malonaldehyde molecules. We discuss the performance of our ML algorithm as well as use it, together with MD simulations, to simulate Raman spectra of the considered systems. The MD simulations are performed using MACE-OFF23,\cite{kovacs2023mace} a recently developed transferable ML force field. This also gives us an opportunity to asses the veracity of this force field in the considered test cases. Our simulated spectra are shown to agree very well with the experiments simultaneously  demonstrating the effectiveness of our ML algorithm as well as the accuracy of the MACE-OFF23 potential.

\section{Methodology \label{dology}}
\subsection{Bond polarizability model (BPM)}
Within the conventional bond polarizability model,\cite{cardona1982light,wolkenstein} the total molecular polarizability, $\boldsymbol{\alpha}$, is given by the sum of bond polarizabilities, $\boldsymbol{\alpha}^b$,
\begin{equation}\label{BPM1}
   \boldsymbol{\alpha} = \sum_b  \boldsymbol{\alpha}^b
\end{equation}
The elements of the individual bond polarizability tensors are given by
\begin{equation}\label{BPM2}
    \alpha^b_{ij} = \dfrac{1}{3}\left(2 \alpha_p^b + \alpha_l^b \right) \: \delta_{ij} + \left(\alpha_l^b - \alpha_p^b \right)  \left( \hat{R}_i^b \hat{R}_j^b - \dfrac{1}{3}  \delta_{ij} \right)  ~,
\end{equation}
where $\alpha_l^b$ and $\alpha_p^b$ are the longitudinal and perpendicular polarizabilities of bond $b$, respectively, $\hat{\textbf{R}}^b$ is a unit vector along the bond $b$, and $i,j=\{x,y,z\}$.
Finally, within the conventional implementation of BPM, the parameters $\alpha_l^b$ and $\alpha_p^b$ are assumed to vary only with the length of the bond $b$. 

It is important to note that the bond polarizability model assumes that the considered bonds are cylindrically symmetric. 
Although this assumption is not strictly valid except in most symmetric systems, BPM has been successfully used in a range of solid state materials.\cite{cardona1982light,mazzarello2010signature,umari2001raman,berger2024polarizability,liang2017interlayer}
It is also possible to apply BPM in molecular systems.\cite{hermet2006raman,paul2024accuracy} However, as was very recently demonstrated by Paul \textit{et al.},\cite{paul2024accuracy}  the conventional BPM generally lacks the desired quantitative accuracy especially in the case of flexible and low-symmetry systems. The most likely reason for the failure of BPM is thought to be the assumption that the bond polarizabilties, and therefore also the total polarizability, depend only on the length of the considered bonds.
We removed this constraint in our algorithm, in addition to several technical differences in the calculations.

\subsection{Machine learning model \label{KRR}}
In this Section, we demonstrate how a KRR algorithm can be built based upon the BPM discussed above. 
We first note that the molecular polarizability tensor $\boldsymbol{\alpha}$ can be conveniently separated into its isotropic and anisotropic components,
\begin{equation}
    \boldsymbol{\alpha} = \alpha_{iso} \mathbb{1} + \boldsymbol{\beta},
\end{equation}
where the isotropic polarizability is $\alpha_{iso} = (\alpha_{xx} + \alpha_{yy} + \alpha_{zz})/3$, and $\boldsymbol{\beta}$ is the (symmetric and traceless) anisotropic polarizability tensor.
It is then useful to re-write Eq.~\eqref{BPM2} for the components of the bond polarizability tensor as 
\begin{equation}\label{BPML}
    \alpha^b_{ij} = \alpha_\mathrm{iso}^b \: \delta_{ij} + \beta^b \left( \hat{R}_i^b \hat{R}_j^b - \dfrac{1}{3}  \delta_{ij} \right)  ~,
\end{equation}
where we define the isotropic and anisotropic bond polarizabilities as $\alpha_\mathrm{iso}^b = (2\alpha_p^b + \alpha_l^b)/3$ and $\beta^b = \alpha_l^b - \alpha_p^b$, respectively. 

The machine learning task is therefore reduced  to inferring the isotropic and anisotropic parameters, $\alpha_\mathrm{iso}^b$ and $\beta^b$, for each of the considered bonds depending on their character, length, and wider nuclear environment. We assume that all these bond characteristics can be encoded within feature vectors denoted as $\textbf{q}^b$.
We propose to use rotationally and translationally invariant descriptors, such as the atom-centered symmetry functions (ACSF) or  smooth overlap of atomic
positions (SOAP) descriptors, as used in ML force fields which model the total molecular energy as a sum of atomic energies.\cite{behler2011atom,christensen2020fchl,de2016comparing} Here, however, the descriptors are evaluated at bond-center rather than atomic positions. 
It is also possible to obtain bond descriptors as a sum of the two relevant atom-centered descriptors. As we show in Section S1, doing so can yield models with similar efficacy to the approach pursued here. 

As can be seen from Eq.~\eqref{BPML}, the isotropic and anisotropic components are readily determined separately.
The (scalar) isotropic component can be evaluated using conventional kernel ridge regression, formally identical to KRR models used to machine-learn molecular energies\cite{huang2020quantum,bartok2015g,christensen2020fchl} although, as discussed above, for consistency we consider here a sum over bond rather than atomic contributions. Namely,
\begin{equation}\label{isoeq}
    \alpha_{iso} = \sum_b \alpha_{iso}^b = \sum_n\sum_{b, b'} K(\textbf{q}^b,\textbf{q}^{b'}) \: w_n ~,
\end{equation}
where the sum over $n$ runs over the training compounds, each with bonds $b'$, and $w_n$ are the regression coefficients. $K(\textbf{q}^b,\textbf{q}^{b'})$ is the kernel which compares the specified feature vectors. Throughout this work, we will make use of the Gaussian kernel 
\begin{equation}
    K(\textbf{q}^b,\textbf{q}^{b'}) = \exp\left( - \gamma \: || \textbf{q}^b - \textbf{q}^{b'}||^2\right) ~,
\end{equation}
where $\gamma$ is one of the hyperparameters of the algorithm.

For a set of geometries, Eq.~\eqref{isoeq} can be vectorized as $\boldsymbol{\alpha}_{iso} = \textbf{K}\textbf{w}$. The solution for the coefficient vector \textbf{w} is found by minimizing the relevant cost function on a set of training geometries.\cite{rupp2012fast} It is given by
\begin{equation}\label{www}
    \textbf{w} = \left( \textbf{K} + \lambda \mathbb{1} \right)^{-1} \boldsymbol{\alpha}_{iso} ~,
\end{equation}
where $\lambda$ is the regularization constant which constitutes the second hyperparameter of the algorithm.

An analogous model can be set up for each of the components of the anisotropic polarizability tensor. Based on Eq.~\eqref{BPML}, they can be evaluated as
\begin{equation}\label{beta}
    \beta_{ij} = \sum_b \beta^b Q_{ij}^b = \sum_n\sum_{b,b'} K(\textbf{q}^b,\textbf{q}^{b'}) \: Q_{ij}^b \: v_n ~,
\end{equation}
where for brevity we define $Q_{ij}^b =  \hat{R}_i^b \hat{R}_j^b - \dfrac{1}{3} \:  \delta_{ij}$, and $v_n$ are the anisotropic regression coefficients which are identical for all of the components $\beta_{ij}$.
This expression can be similarly vectorised as $\boldsymbol{\beta}_{ij} = \textbf{K}_{ij} \textbf{v}$ although we note the differences between the $\textbf{K}$ and $\textbf{K}_{ij}$ kernel matrices.
It is possible to determine the regression coefficient $v_n$ by considering a single set of components $\boldsymbol{\beta}_{ij}$ and using an expression analogous to Eq.~\eqref{www}. Doing so, however, generally delivers sub-optimal performance since only a fraction of the training data is utilised.

Instead, we will consider the following loss function,
\begin{equation}
    \mathcal{L} = \dfrac{1}{2}\sum_{i,j} || \boldsymbol{\beta}_{ij} - \textbf{K}_{ij} \textbf{v}||^2 ~.
\end{equation}
The coefficient vector $\textbf{v}$ can be found by setting $\mathrm{d}\mathcal{L}/\mathrm{d}\textbf{v} = 0$. It is hence given by the solution to
\begin{equation}\label{solv}
    \left(\sum_{i,j} \textbf{K}_{ij} \textbf{K}_{ij} \right) \textbf{v} = \sum_{k,l} \textbf{K}_{kl} \boldsymbol{\beta}_{kl} ~.
\end{equation}
To avoid over-fitting, the vector $\textbf{v}$ in Eq.~\eqref{solv} above will be determined using singular value decomposition in which the singular values of the matrix on the left-hand-side will be disregarded if below a set threshold. This threshold, $\epsilon$, can be therefore regarded as a de facto regularization parameter.\cite{christensen2019operators}

We finish with several remarks regarding the method outlined here. Firstly, we note that the (scalar) isotropic polarizability could be in principle evaluated using any number of other algorithms. It is possible, for instance, to assume that the isotropic polarizability is given by a sum of atomic, rather than bond, contributions or indeed use any of the methods developed for machine learning of potential energy or other scalar quantities.
Our choice here was driven purely by the desire for consistency with the algorithm for the anisotropic part of the polarizability tensor.
Nonetheless, in Section S1.1, we compare the modelling of the isotropic polarizability as a sum over atomic and bond contributions and find these two approaches to have similar accuracy.

Secondly, as can be inferred from Eq.~\eqref{beta}, by its construction our algorithm ensures that the anisotropic polarizability matrix $\boldsymbol{\beta}$ is traceless and symmetric, as desired. 
Thirdly, one can see that the evaluation of the anisotropic tensor components, using Eq.~\eqref{beta}, does not involve any significant additional computational cost compared to the isotropic component which is evaluated using Eq.~\eqref{isoeq} since the computationally most intensive part of Eq.~\eqref{beta} is determining the similarity kernel $K(\textbf{q}^b,\textbf{q}^{b'})$.   
Lastly, the \textit{bonds} considered within our algorithm are defined simply by the relevant pairs of atoms. It is therefore possible, and important in some cases, to introduce fictitious bonds into the model which will capture the relevant non-covalent interactions or allow for, for instance, applying this model to systems undergoing bond breaking/forming processes; see the example of malonaldehyde below.

\subsection{Raman spectra \label{raman}}
It is possible to combine ML potential and dipole or polarizability algorithms to evaluate the harmonic IR or Raman molecular spectra, respectively. This was very recently demonstrated by Pracht \textit{et al.} on the example of IR spectroscopy also using the MACE-OFF23 force field.\cite{pracht2024efficient}
Here, however, in order to capture anharmonic effects as well as a possible influence of conformational changes on the Raman spectrum, we refrain from calculating the spectra through harmonic analyses.
Instead, they are evaluated by considering the relevant polarizability correlation functions obtained from molecular dynamics trajectories using the ML algorithm outlined above.

As is well established,\cite{raman2002long,thomas2013computing} the isotropic part of the  Raman spectrum can be calculated from the derivatives of the polarizability tensor components as
\begin{equation}\label{iso}
    I_{iso}(\omega) \propto \nu(\omega) \int \mathrm{d}t \:  e^{\mathrm{i}\omega t} \:\langle \dot{\alpha}_{iso}(\tau) \dot{\alpha}_{iso}(t-\tau)  \rangle_\tau
\end{equation}
The anisotropic part of the spectrum can be similarly written as 
\begin{equation}\label{aniso}
    I_{aniso}(\omega) \propto \nu(\omega) \int \mathrm{d}t \:  e^{\mathrm{i}\omega t} \: \langle \mathrm{Tr}\left[\dot{\boldsymbol{\beta}}(\tau) \dot{\boldsymbol{\beta}}(t-\tau)\right]  \rangle_\tau
\end{equation}
where $\langle \cdot \rangle_\tau$ denotes the ensemble average over $\tau$, and the prefactor $\nu(\omega)$ is
\begin{equation}
    \nu(\omega) = \dfrac{(\omega - \omega_0)^4}{\omega } \dfrac{1}{1 - \exp(-\hbar\omega/k_\mathrm{B}T)}
\end{equation}
where $\omega_0$ is the frequency of incident light.
Finally, the total spectrum is then given by\cite{raman2002long,brehm2019computing}
\begin{equation}
    I(\omega) \propto I_{iso}(\omega) + \dfrac{7}{30} I_{aniso}(\omega)
\end{equation}

\section{Computational details \label{methods}}
In the subsequent section, we make use of the polarizability algorithm described in Section \ref{KRR} and the formulae discussed in Section \ref{raman} to simulate Raman spectra of two molecular systems: biphenyl and malonaldehyde. Here, we briefly describe the details of the electronic structure calculations, MD simulations, and ML algorithm training.\\
\textbf{Electronic structure calculations:}
The static polarizability tensors used in the training of the ML algorithms were calculated using NWChem 7.0.0.\cite{apra2020nwchem,autschbach2011time} with the B3LYP functional and cc-pVTZ or aug-cc-pVTZ basis set for biphenyl and malonaldehyde, respectively. The polarizability values are reported in atomic units.\\
\textbf{ML algorithm and training:}
Throughout this work, in the kernel polarizability algorithm, we use the Smooth Overlap of Atomic Positions (SOAP) descriptors,\cite{de2016comparing} as implemented in the DScribe package,\cite{himanen2020dscribe} evaluated at the center of each of the considered bonds. 
We use a cut-off of 6 \AA, and set $n_{max} = 6$ and $l_{max} =4$.
In the case of biphenyl, only the twenty-three covalent bonds are considered. For malonaldehyde, in addition to all covalent bonds, both of the possible O-H bonds are included. The hyper-parameters of the models were optimized by cross-validation. For both ML algorithms, we obtained and use $\gamma = 1 \times 10^{-5}$, $\lambda = \epsilon = 1 \times 10^{-8}$.\\
\textbf{MD simulations:}
The MD simulations used to obtain the polarizability correlation functions were performed using (the `medium' variant of) the recently developed MACE-OFF23 force field\cite{kovacs2023mace} interfaced with the Atomic Simulation Environment (ASE).\cite{larsen2017atomic} 
The MD simulations are run using a time step of 1 fs throughout.
In all cases, the system is equilibrated for 50 ps which is followed by a 1 ns NVT simulation during which the polarizability tensor is evaluated at every step. For biphenyl, the simulation is run at the experimental temperature of 363 K. For malonaldehyde, the simulations are performed at three different temperatures: 200 K, 30 K, and 16 K. 
We make use of the Langevin thermostat with the damping constant of $\gamma_D = 1$ ps$^{-1}$ during the equilibrations and during the biphenyl simulation run. A damping constant of $\gamma_D = 0.1$ ps$^{-1}$ is used during malonaldehyde simulation runs. \\
\textbf{Raman spectra:}
The Fourier transforms in Eqs.~\eqref{iso} and \eqref{aniso} are evaluated using Burg's maximum entropy method\cite{burg1975maximum} as implemented in the memspectrum package.\cite{memspectru} The frequencies are reported with no frequency scaling often applied to compensate for approximations involved in the electronic structure calculations.
Following the relevant experimental studies,\cite{carreira1977raman,luttschwager2013vibrational} the wavelength of incident light was set to 488 and 532 nm in the case of biphenyl and malonaldehyde spectra, respectively.

\section{Applications \label{results}}

\subsection{Biphenyl}

\begin{figure*}
    \centering
    \includegraphics{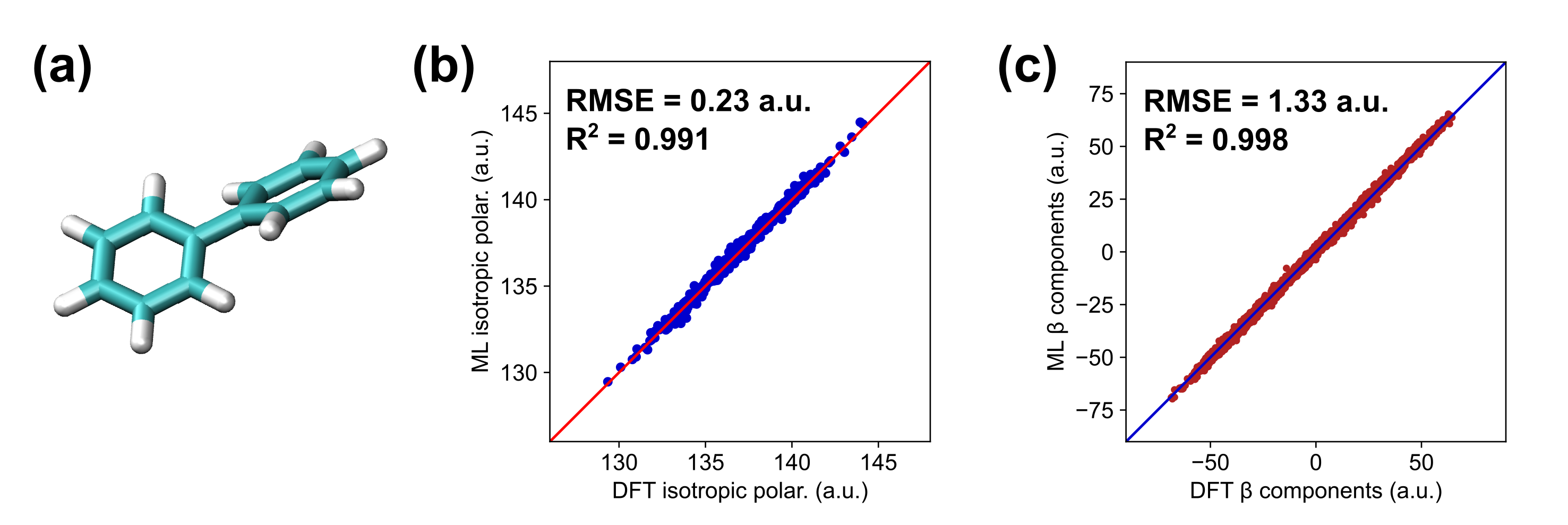}
    \caption{(a) Schematic structure of the biphenyl molecule. (b, c) Correlations between the DFT and ML values of (b) the isotropic polarizability, $\alpha_{iso}$, and (c) unique elements of the anisotropic polarizability tensor, $\beta_{ij}$.}
    \label{bph1}
\end{figure*}

Biphenyl is a flexible molecule with a non-planar energy minimum geometry in gas phase,\cite{barrett1972vibrational} see Figure \ref{bph1}(a). Its experimental Raman spectrum, down to the range of approximately 30 cm$^{-1}$, has been previously reported in Ref.~\onlinecite{carreira1977raman} making it a well-suited benchmark system for our purposes.

We begin by developing a ML polarizability algorithm following the framework described in Section \ref{KRR}. The training geometries are taken from a classical MD simulation performed at 500 K to ensure good sampling of molecular geometries while the testing geometries were drawn from the 50 ps equilibration trajectory (at the experimental temperature of 363 K). The remaining details of the training procedure are described in Section \ref{methods}.
The performance of the isotropic and anisotropic algorithms, following their training on the same set of 600 molecular geometries, is shown in Figure \ref{bph1}(b) and (c), respectively. Overall, we obtain good performance on the testing set. For the isotropic polarizability $\alpha_{iso}$, we find the root-mean-squared error (RMSE) of 0.23 a.u., relative to the standard deviation within the testing dataset of 2.42 a.u., and the value of $R^2 = 0.991$.
For the anisotropic components $\beta_{ij}$, the algorithm yields the RMS error of 1.33 a.u., relative to the testing set standard deviation of 28.6 a.u., and $R^2 = 0.998$.
It is possible to reduce the errors further by incorporating more training data. As we demonstrate in Section S2.1, however, only relatively modest improvements in the algorithms’ performance can be achieved by doing so. Furthermore, as is typical in the case of kernel-based algorithms, this is associated with an increased prediction computational cost and, as we will demonstrate below, the present accuracy is sufficient to very well reproduce the experimental signal.

\begin{figure*}
    \centering
    \includegraphics{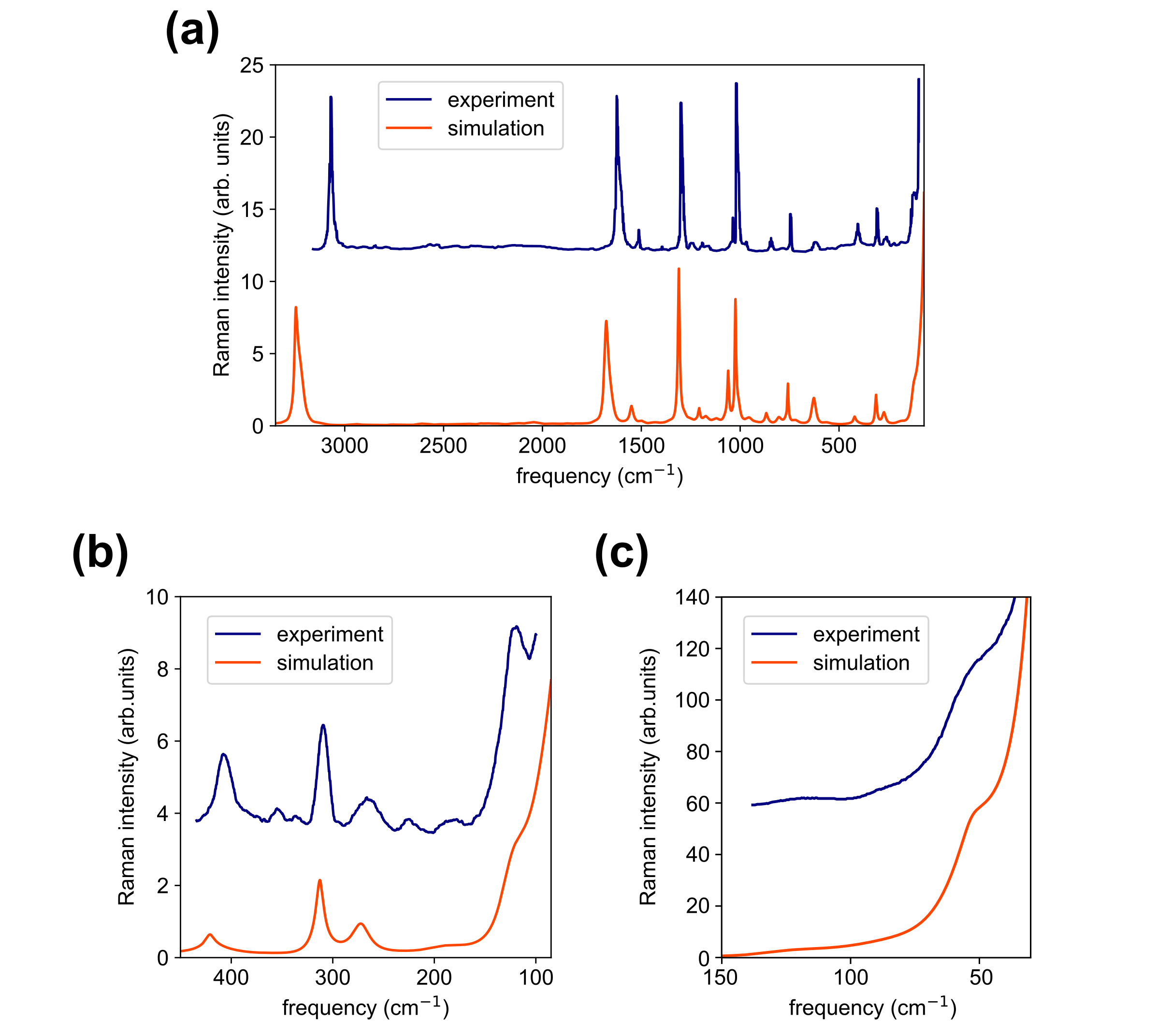}
    \caption{Simulated and experimental Raman spectrum of biphenyl in gas phase over (a) the entire frequency range and (b, c) in the low frequency regions. Experimental data has been extracted from figures in Ref.~\onlinecite{carreira1977raman} and off-set vertically from the simulation results for easier comparison. }
    \label{bph2}
\end{figure*}

We next proceed to evaluate the Raman spectrum of the considered system at the experimental temperature of 363 K. As discussed in detail in Section \ref{methods}, the spectrum is obtained from the autocorrelation functions of the polarizability tensor components obtained from the MD trajectory. 
The resulting spectrum over the entire frequency, accompanied by the experimental signal from Ref.~\onlinecite{carreira1977raman}, is shown in Figure \ref{bph2}(a). The simulated spectrum agrees very well with the experiment both in terms of peak positions as well as their line shapes. We find, for instance, a single slightly asymmetric peak above 3000 cm$^{-1}$ which corresponds to the C-H bond stretches, as well as a pronounced peak around 1600 cm$^{-1}$ stemming from the ring stretching mode. Similarly, we find very good agreement between the simulation and experiment in the fingerprint region below 1500 cm$^{-1}$. 
The peak frequencies tend to be somewhat overestimated in the simulated spectrum as is typical when using electronic structure method to calculate vibrational spectra. This issue is usually addressed by introducing an empirical frequency scaling factor.\cite{scott1996harmonic,kesharwani2015frequency} In the present case a scaling factor of around 0.98 would result in an improved frequency match with the experiment but, as discussed in the Methods section, we refrained from introducing such scaling here. 

Figures \ref{bph2}(b) and (c) show the low frequency regions of the spectrum. We again find very good agreement between the simulated and experimental spectra. In particular, we can see a relatively pronounced shoulder at around 55 cm$^{-1}$ in both the simulation and the experiment which almost certainly corresponds to the biphenyl torsional mode. As in the experiment, this mode is also present in the isotropic (polarized) part of the spectrum, see Figure S4.
Finally, as shown in Figure S5, we note that the 110 cm$^{-1}$ mode becomes more pronounced in the simulation in the case of slower thermostatic damping, in better agreement with the measurement.
%
%
%There are however two very weak peaks in the experimental spectrum, at around 185 and 222 cm${-1}$, that are missing from the calculated Raman signal.
%To resolve this issue, we performed harmonic analyses on the biphenyl molecule using several electronic structure methods, see Section SX. None of these calculations find any vibrational modes in the range of 150-260 cm$^{-1}$, in agreement with our MD simulations. This suggest that the very weak 185 and 222 cm${-1}$ peaks most likely correspond to vibrational overtones. 
%
%
In summary, therefore, the ML polarizability algorithm developed here, together with the MACE-OFF23 force field, yields a Raman spectrum that is in a very good agreement with the experiment.
%Not only does it underscore the efficacy of our bond polarizability model but also demonstrates the accuracy of the used transferable ML force field in the present setting. 

\subsection{Malonaldehyde}
\begin{figure*}
   \centering
    \includegraphics{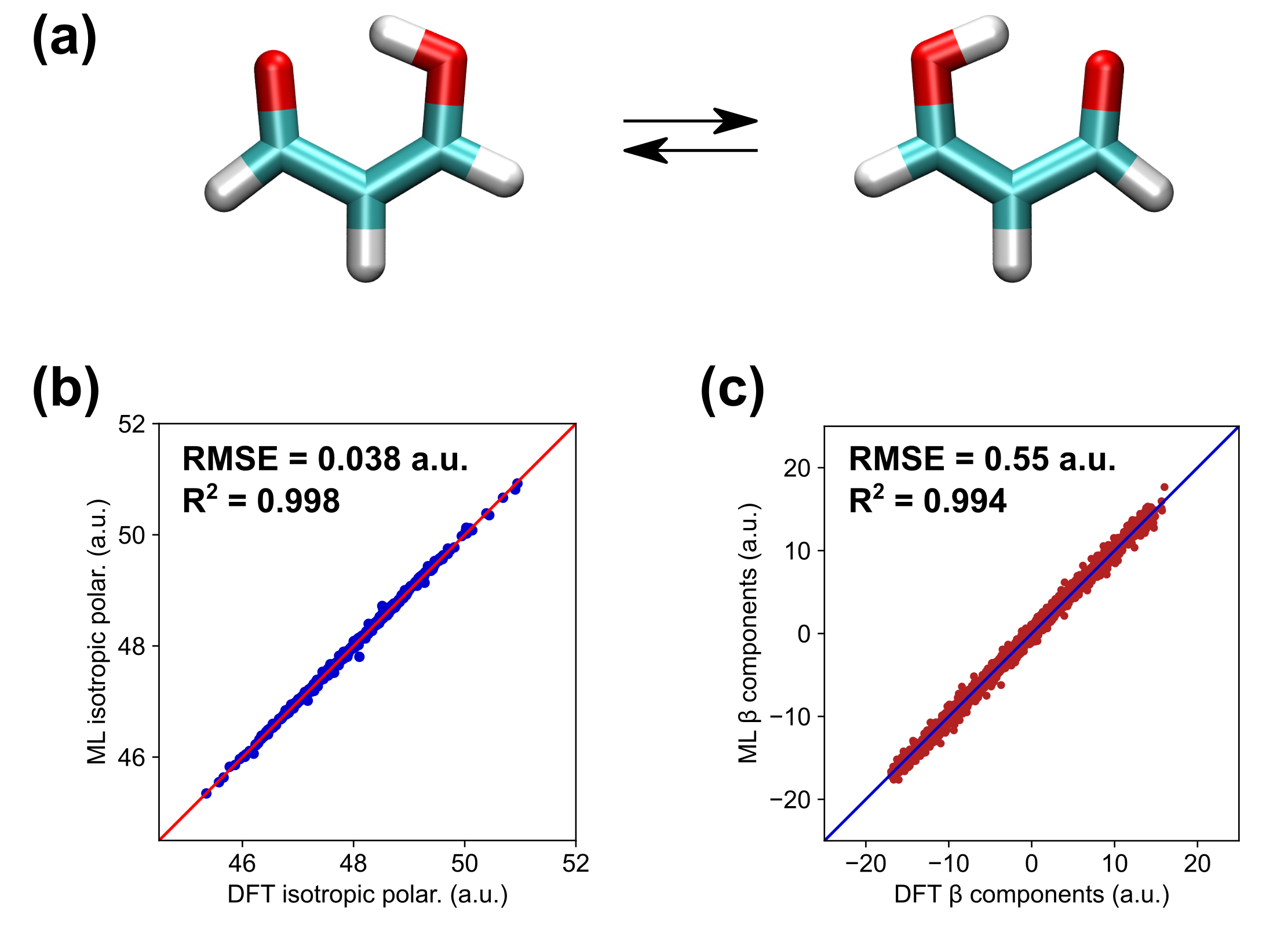}
    \caption{(a) Schematic structures and proton transfer reaction of malonaldehyde. (b, c) Correlations between the DFT and ML values of (b) the isotropic polarizability, $\alpha_{iso}$, and (c) unique elements of the anisotropic polarizability tensor, $\beta_{ij}$.}
    \label{mal1}
\end{figure*}

We next move on to consider malonaldehyde (MA). MA can exist in keto and enol forms of which the enol form (considered here) is known to be more stable in gas phase.\cite{yamabe2004reaction} As shown in Figure \ref{mal1}(a), (enol) MA is known to undergo an intra-molecular proton transfer, and has long served as a served as a model system for this process.\cite{caldin2013proton} Consequently, the MA dynamics as well as its IR spectrum have attracted sustained interest over the years.\cite{hammer2011intramolecular,chiavassa1992experimental,firth1989matrix,smith1983infrared,hazra2009combining,yang2010generalized} 
Notably in our context, the IR spectrum of malonaldehyde has been also studied using ML force field and dipole algorithms by K\"{a}ser \textit{et al.}\cite{kaser2020reactive, kaser2023transfer} On the other hand, the experimental MA Raman spectrum has been reported by L\"{u}ttschwager \textit{et al.} only relatively recently\cite{luttschwager2013vibrational,luttschwager2010periodic} and has not yet received comparable attention.

As in the case of the previous test system, we begin by training the ML polarizability models. We use 600 geometries extracted from a 500 K MD simulation, once again in order to ensure sufficient sampling of the phase space. 
The performance of the isotropic and anisotropic algorithms on the testing set, which in this case was drawn from an independent 500 K simulation, is shown in Figures \ref{mal1}(b) and (c). For the isotropic part, we find RMSE of 0.038 a.u., as compared to the set standard deviation of 0.92 a.u., and $R^2 = 0.998$. For the anisotropic algorithm, we obtain RMSE of 0.55 a.u., relative to the testing set standard deviation of 7.49 a.u., and $R^2 = 0.994$.

Since the proton transfer dynamics within malonaldehyde has been examined repeatedly in other works,\cite{kaser2020reactive,kaser2023transfer,wang2008full,yang2010generalized} we refrain from discussing it in detail here. We note however that we find that according to the MACE-OFF23 potential, the barrier  for proton transfer is approximately 5.2 kcal/mol, see Section S3. This is in relatively good agreement with the CCSD(T) result of Wang \textit{et al.} who obtained a value of 4.1 kcal/mol\cite{wang2008full} and results in a relatively infrequent proton transfer in the present classical MD simulations  below room temperature. 

Instead, we focus on the MA Raman spectrum. We note that due to proton tunneling between the potential wells corresponding to the two isomers shown in Figure \ref{mal1}, splitting within the spectroscopic vibrational peaks is generally experimentally observed.\cite{luttschwager2010periodic}
These tunnel splittings vary between different modes but cannot be recovered in our classical MD-based simulations of the Raman signal. 
They can be evaluated with an ML potential for instance using the ring-polymer instanton method,\cite{richardson2018perspective} as has been recently demonstrated by Käser et al.\cite{kaser2022transfer} Calculating a Raman spectrum which captures the tunnel splittings directly may similarly be possible if the polarizability correlation functions are evaluated using path integral-based approximate quantum dynamics methods, such as ring polymer molecular dynamics,\cite{habershon2013ring} the so-called “Matsubara dynamics”,\cite{hele2015boltzmann,smith2015new} or the more general approach developed by Hammes-Schiffer and co-workers,\cite{chow2023nuclear} rather than purely classical MD simulations as is done here.     
Nonetheless, for most of the vibrational modes, the tunnel splittings are sufficiently modest to enable a meaningful comparison between the experimental spectra from Ref.~\onlinecite{luttschwager2013vibrational} and the simulated ones evaluated using the MACE-OFF23 potential and our ML polarizability algorithm.
Here, we consider the spectrum measured at sub-ambient temperature while the spectra evaluated and measured at 16 and 30 K are discussed in Section S4.2. 

\begin{figure}
    \centering
    \includegraphics{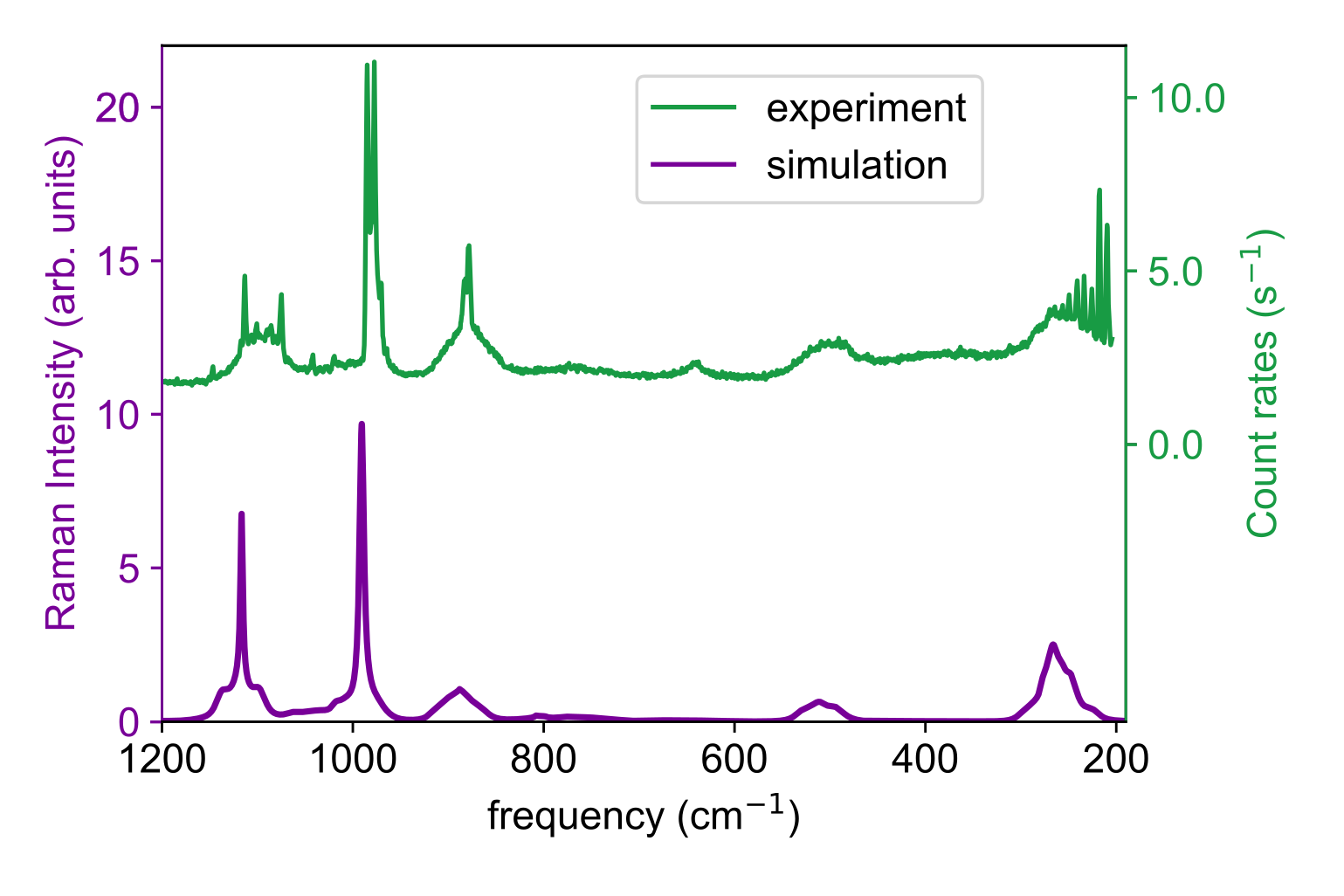}
    \caption{Simulated and experimental Raman spectrum of malonaldehyde. In the experimental spectrum, the sharp lines below 400 cm$^{-1}$ arise primarily from residual HCl, see Ref.~\onlinecite{luttschwager2013vibrational} for detailed discussion.}
    \label{mal2}
\end{figure}
In Figure \ref{mal2}, we show the comparison between the experimental Raman signal and the simulated spectrum evaluated at 200 K. 
We focus here on the lower frequency range due to the presence of a handful of unassigned peaks above roughly 1200 cm$^{-1}$ in the experiment.\cite{luttschwager2013vibrational}  For completeness, the Raman spectra over the entire range are plotted in Figure S7.
We find that the two spectra in Figure \ref{mal2} are in very good agreement although, as discussed above, the simulation does not capture the tunneling effects. 
In agreement with our discussion of the biphenyl spectra, the peak frequencies are somewhat overestimated due to the absence of an empirical scaling factor in our calculations.
Nonetheless, the simulation predicts the structure of the spectrum well including the presence of the roughly 250 and 500 cm$^{-1}$ peaks corresponding to collective bends; see Ref.~\onlinecite{luttschwager2013vibrational} for a detailed discussion of mode assignment.
Similarly, we find a very intense (C-C stretch) mode around 1000 cm$^{-1}$ as well as those of in-plane ring and C-H bending at around 880 and 1100 cm$^{-1}$, respectively.
The simulated spectrum also features a very weak peak at around 670 cm$^{-1}$ although its intensity appears to be underestimated as compared to the experiment. 
The vibrational peak corresponding to the O-O stretching vibration, which is known to act as a gating mode in the proton transfer dynamics, is found in the simulations at around 280 cm$^{-1}$ although it overlaps with the approximately 250 cm$^{-1}$ out-of-plane bending mode. Nonetheless, it can be easily identified by its presence in the polarised (isotropic) spectrum, see Figure S8, and is very clearly resolved in the lower temperature spectra, see Figure S9. 
In the Raman experiment, this peak is split into peaks at roughly 240 and 480 cm$^{-1}$ precluding a straightforward comparison. 
%
%The polarised spectrum 

We similarly find that the measured and simulated spectra agree with each other well at lower temperatures, see Section S4.2. 
Overall, although our simulations do not account for tunnel splitting effects, the ML polarizability algorithm together with the MACE-OFF23 potential again proved able to reproduce the experimental Raman spectra well, underscoring the robustness of both the polarizability algorithm as well as the considered force field.

\section{Conclusions}
In this work, we introduced a KRR machine learning algorithm based on the bond polarizability model which can be used to predict the molecular electric polarizability tensor. Due to its architecture, the model is guaranteed to yield rotationally equivariant tensors at a computational cost and implementation complexity similar to that of scalar kernel ridge regression. We have subsequently applied this algorithm to machine learning the polarizability tensor of two model systems, biphenyl and malonaldehyde. The ML models were shown to perform well for both the isotropic and anisotropic tensor components.
We then used them, together with classical MD simulations run with the MACE-OFF23 potential, to calculate the Raman signal of the considered molecular species. The resulting spectra agreed very well with the respective experimental measurements demonstrating not just the utility of our ML algorithm but also the accuracy of this recently developed transferable ML force field.

Although the algorithm developed here was based on the KRR framework, it is relatively straightforward to build an analogous deep neural network implementation of BPM. Additionally, it may also be possible to extend the algorithm presented here to go beyond including only the pre-defined bonds. Instead, one could consider all atomic pairs within a chosen distance with a smooth cut-off function. Such a modification of our approach can be expected to increase the accuracy of the ML model although likely at the cost of increased computational intensity of both the training and the evaluation tasks.

Overall, this work has demonstrated the utility of physics-based approaches to deliver relatively simple but effective ML algorithms for molecular Raman spectra and showcased how such models can be benchmarked against experiment.

%%%%%%%%%%%%%%%%%%%%%%%%%%%%%%%%%%%%%%%%%%%%%%%%%%%%%%%%%%%%%%%%%%%%%
%% The "Acknowledgement" section can be given in all manuscript
%% classes.  This should be given within the "acknowledgement"
%% environment, which will make the correct section or running title.
%%%%%%%%%%%%%%%%%%%%%%%%%%%%%%%%%%%%%%%%%%%%%%%%%%%%%%%%%%%%%%%%%%%%%
\section*{Acknowledgement}
This work was supported by the Center for Adopting Flaws as Features, an NSF Center for Chemical Innovation supported by grant CHE-2413590. We also acknowledge the Big-Data Private-Cloud Research Cyberinfrastructure MRI award funded by the NSF under grant CNS-1338099 and by Rice University’s Center for Research Computing (CRC). 
We thank Dr Nils L\"{u}ttschwager and Prof Martin Suhm for sharing their original experimental data reported in Ref.~\onlinecite{luttschwager2013vibrational}.

%%%%%%%%%%%%%%%%%%%%%%%%%%%%%%%%%%%%%%%%%%%%%%%%%%%%%%%%%%%%%%%%%%%%%
%% The same is true for Supporting Information, which should use the
%% suppinfo environment.
%%%%%%%%%%%%%%%%%%%%%%%%%%%%%%%%%%%%%%%%%%%%%%%%%%%%%%%%%%%%%%%%%%%%%

\section*{Data availability statement}
\noindent A Python implementation of the ML algorithm, an example of a machine learning workflow as well as the DFT data used in this study are openly available at https://github.com/jakubks/BPM

\end{document}